\documentclass[aps,prb,twocolumn,amssymb]{revtex4}

\usepackage{epsfig}

\begin{document}
\renewcommand{\thefootnote}{\fnsymbol{footnote}} 
\renewcommand{\theequation}{\arabic{section}.\arabic{equation}}

\title{Controlling crystallization and its absence: proteins, colloids and
patchy models}

\author{Jonathan P.\ K.\ Doye,$^a$  Ard A.\ Louis,$^b$  I-Chun Lin,$^c$ 
Lucy R.\ Allen,$^d$  Eva G.\ Noya,$^e$  Alex W.\ Wilber,$^a$  
Hoong Chwan Kok,$^e$  Rosie Lyus,$^e$}

\affiliation{$^a$ Physical and Theoretical Chemistry Laboratory,
Oxford University, South Parks Road, Oxford OX1 3QZ, United Kingdom\\
$^b$ Rudolf Peierls Centre for Theoretical Physics, 1 Keble Road,
Oxford OX1 3NP, United Kingdom\\
$^c$  Institute of Chemical Sciences and Engineering,
BCH 4109 Swiss Federal Institute of Technology EPFL,
CH-1015 Lausanne, Switzerland\\
$^d$ School of Physics and Astronomy, University of Leeds LS2 9JT, 
United Kingdom\\
$^e$ Department of Chemistry, University of Cambridge,
Lensfield Road, Cambridge, CB2 1EW, United Kingdom}

\date{\today}

\begin{abstract}
\noindent The ability to control the crystallization behaviour (including
its absence) of particles, be they biomolecules such as globular proteins,
inorganic colloids, nanoparticles, or metal atoms in an alloy,
is of both fundamental and technological importance.
Much can be learnt from the exquisite control that biological systems 
exert over the behaviour of proteins, where protein crystallization and
aggregation are generally suppressed, but where in particular instances complex
crystalline assemblies can be formed that have a functional purpose. 
We also explore the insights that can be obtained from computational modelling,
focussing on the subtle interplay between the interparticle interactions, the
preferred local order and the resulting crystallization kinetics. 
In particular, we highlight the role played by ``frustration'', where 
there is an incompatibility between the preferred local order and the global
crystalline order, using examples from atomic glass formers and 
model anisotropic particles.
\end{abstract}

\maketitle

\section{Introduction}
\label{sect:intro}
Controlling crystallization is a subject of considerable importance both from
a fundamental and an applied perspective.
Chemical physicists want to understand how the interparticle interactions 
determine the most favoured crystal structure and the ease with which 
crystallization can occur.
Biochemists want to know the best recipe to crystallize the protein in which
they are interested, so that they can then determine its structure.
Nanotechnologists want to know how to get colloids or nanoparticles
to self-assemble into a given target structure, such as a diamond
lattice because of its potential importance in photonics.
Metallurgists want to be able to predict which alloys will most readily avoid
crystallization, and instead form a metallic glass.\cite{Miracle04} 

In this paper we want to take a theoretical and computer simulation perspective 
on the factors that control crystallization, and its absence, in these kinds 
of systems.  So far in the literature, isotropic models have been the starting 
point for much of the theoretical work in this area, and although there have 
been considerable successes, this approach has its limits.
For example, the experimentally observed ``crystallization slot''
for globular proteins, where the interprotein interactions,
as measured by the second virial coefficient, are sufficiently
attractive to encourage crystallization, but not so attractive as to lead
to irreversible aggregation,\cite{George94}  has been rationalised using 
isotropic model potentials.\cite{Rosenbaum96}
Another fruitful idea is that 
if a protein solution lies near to a metastable critical point associated
with separation into protein-rich and solvent-rich phases,
the associated concentration fluctuations could enhance crystal 
nucleation.\cite{tenWolde97}
However, proteins are not isotropic, but are anisotropic both in their
shape and in their interactions. One reflection of the latter is that 
protein crystals typically have much lower packing fractions\cite{Matthews68} 
than the close-packed structures that are favoured by isotropic 
potentials.

Similarly, for colloids and nanoparticles there are now a considerable array of
different crystals that have been obtained just using
isotropic particles, particularly through the use of binary mixtures. 
For example, just by varying the relative sizes of 
the two types of particles, surprisingly complex crystals can be 
formed.\cite{Bartlett92,Eldridge93a} Very recently, the use of particles
with charges of opposite signs has led to colloidal\cite{Bartlett05,Leunissen05}
and nanoparticle \cite{Shevchenko06,Frenkel06,Kalsin06} analogues of ionic 
crystals, even relatively low-density structures such as zinc 
blende.\cite{Kalsin06} 
However, there are limits to the structures that can be assembled from
isotropic particles,\cite{Likos02} and so there has been considerable recent 
interest in developing methods to produce colloidal particles that 
are anisotropic in shape\cite{Manoharan03,Cho05,vanBlaaderen06} or in their 
interactions.\cite{Jackson04,Cho05,Roh05,Snyder05,Li05b,Roh06}
For all these reasons, there has been much theoretical interest in 
beginning to explore simple anisotropic models and their effects on 
crystallization\cite{Sear99c,Dixit02,Song02,Kern03,Chang04,Chang05,Talanquer05,Zhang06} 
and self-assembly.\cite{Zhang04,Hagan06,vanWorkum06,Wilber06}

Here we provide a particular viewpoint on 
this broad area of controlling crystallization and its absence.
In Section \ref{sect:pxtal} we outline some of our perspectives on protein 
crystallization, particularly why it is important to take into account 
the evolutionary origins of proteins when trying to understand their
crystallization behaviour. In Section \ref{sect:glass} we explore how
the concept of frustration can provide physical insight into the 
structural origins of the glass-forming ability for two of the most
common models for simulating supercooled liquids. 
Finally, in Section \ref{sect:patchy}, we introduce some of our recent 
results that use model patchy particles to explore how the geometry of the 
interparticle interactions can both influence and be used to control
the crystallization and self-assembly behaviour. 

\section{Protein crystallization}
\label{sect:pxtal}
The question concerning protein crystallization that we have been particularly 
seeking to address\cite{Doye04b} is: Why are globular proteins
seemingly so hard to crystallize? 
Although it is not true of all proteins---lysozyme and insulin provide
notable examples of proteins that crystallize relatively easily---there is 
plenty of anecdotal evidence that proteins are often very difficult to
crystallize. 
Furthermore, the rise of structural genomics initiatives, 
has now made it possible to start to quantify this difficulty. 
For example, the success rate for producing X-ray quality crystals has been 
estimated to be roughly 20\% for those prokaryotic proteins that can be 
expressed in soluble form.\cite{Hui03b,Derewenda06}
Moreover, the crystallization approaches used are mainly empirical. For
example, screening kits allow large numbers of solution conditions, which 
have previously been found to be useful for crystallization, 
to be quickly tested.

Our perspective on this question\cite{Doye04b} is that the need for proteins
to function within the crowded cellular environment places major constraints
on the surface properties of proteins. In particular, not only must a protein
interact correctly with its binding partners, but it must also make sure that
it does not stick to anything else in the cell, including other copies of 
itself. 
Failure to do so is likely to be deleterious to the cell, as in the
many protein aggregation diseases, of which the most relevant to this paper 
are those associated with native state aggregation and crystallization.
For example, sickle cell anaemia involves the ordered aggregation of hemoglobin
into fibrils, 
and some forms of anaemia and cataracts are the result of the crystallization
of mutant forms of the haemoglobin\cite{Vekilov02b} and 
$\gamma$-crystallin\cite{Pande01} proteins, respectively.

Thus, our hypothesis is
that the surface properties of proteins have been selected
in order to prevent native-state aggregation and crystallization {\em in vivo} 
(we term this `engineering out' of unwanted properties negative design) 
and that, because of the robustness of the mechanisms used, many proteins
are difficult to crystallize even in the far-from-physiological conditions
explored by the protein crystallographer. Furthermore, there is
a relatively simple way to test this hypothesis.
If negative design is present, then random mutagenesis of surface amino acids
should on average lead to a protein that is easier to crystallize, and the 
two such mutagenesis studies in the literature do indeed find such a 
correlation.\cite{McElroy92,Darcy99}

More useful, though, would be to identify the mechanisms used to achieve this
negative design, since then this raises the possibility of developing
rational strategies to overcome the negative design in order to make 
protein crystallization easier.
Interestingly, bioinformatic analyses have provided a `smoking gun' for 
the potential role of surface lysine residues. 
Lysine is the most common surface amino acid, but yet is the most
underrepresented at functional interfaces\cite{Jones96,LoConte99} 
and at contacts between proteins in crystals.\cite{Dasgupta97}
This of course begs the question: What is it doing there, if it is only
reluctantly involved in interactions? 
One possible answer is that lysine helps to prevent unwanted interactions.

Consistent with this suggestion, the group of Zygmunt Derewenda have shown
that systematically mutating surface lysine residues, 
particularly to alanine,\cite{Longenecker01} 
almost invariably enhances the crystallizability of a 
protein.\cite{Derewenda04,Derewenda06}
The long lysine side-chain has substantial conformational entropy, which 
would be lost if constrained at a protein-protein interface. 
Thus, it has been suggested that lysine stabilizes the free protein in 
solution by providing a `surface entropy shield',\cite{Derewenda06} 
and replacing lysines with the compact alanine is expected to make 
crystallization thermodynamically more favourable.

One possible objection to our negative design hypothesis is that, given that
proteins are irregular objects subject to dynamical fluctuations, it is not 
surprising that they are hard to crystallize. However, further evidence that
the low crystallizability of most proteins is not an intrinsic property, but one
that arises due to selection, comes from instances when protein
crystallization does occur {\em in vivo}, because it achieves some functional
purpose. Nature has no problem getting proteins to crystallize, even in the
`dirty' environment of the cell, when there is a positive selection pressure
for this. 

\begin{figure}[t]
  \begin{center}
   \includegraphics[width=8.5cm]{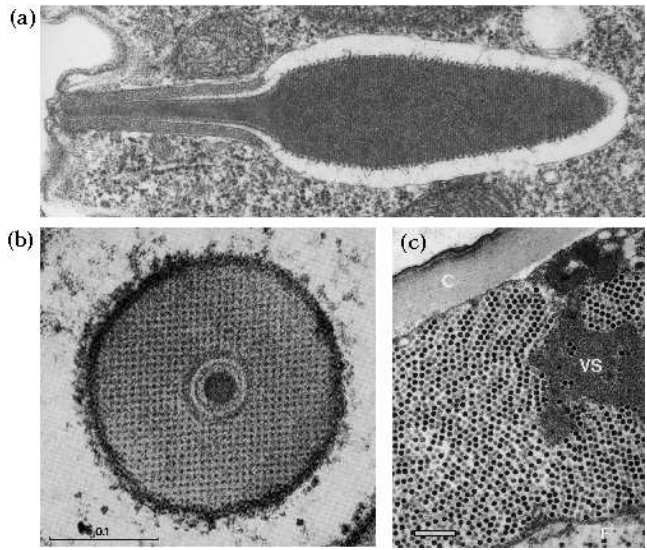}
   \caption{\label{fig:invivo}
    Some examples of protein crystallization occurring in the cell.
    (a) A trichocyst attached to the outer membrane of {\em Paramecium}.
    (b) An encapsulated rod virus of the granulosis virus of {\em Plodia 
        interpunctella}.
    (c) Paracrystalline arrays of an iridescent virus in the epidermis of a 
        larva of {\em Culicoides variipennis sonorensis}.
    The scale bars in (b) and (c) correspond to 0.1$\,\mu$m and 1$\,\mu$m, 
    respectively.
    Reproduced with permission from Refs.\ (a) \onlinecite{Arnott68} and
    (b) \onlinecite{Vayssie00} and (c) \onlinecite{Mullens99}.}
  \end{center}
\end{figure}

There are many fascinating examples of this {\em in vivo} protein
crystallization, as catalogued in a recent review.\cite{Doye06b}
Such crystallization represents a rather neglected area of biological 
self-assembly, and one which is worthy of further study. Here we give just
a few examples. 
Certain genera of viruses that infect insect larvae, 
coopt the cell to express large quantities of specific proteins 
during the late stages of infection. 
These then crystallize around the viral particles to provide a protective
environment for the viruses after the death of the insect larvae, 
as illustrated in Fig.\ \ref{fig:invivo}(b).
On ingestion of the crystals by a new insect larva, the protein crystals
dissolve in the alkaline environment of the gut, releasing the virus 
to infect the new host.\cite{Smith76}

A number of examples of crystalline proteins in the cell correspond to 
proteins stored in secretory granules. Fig.\ \ref{fig:invivo}(a) illustrates 
one particularly specialized example from the protist (a type of 
single-celled eukaryotic organism) {\em Paramecium}.\cite{Vayssie00} 
Large numbers of these granules, which are called trichocysts, 
are attached to the outer membrane of this organism, 
and are assembled from three different 
families of closely related polypeptides, each localized to different 
regions of the trichocyst to form a structure that is of the order of 
3--4$\,\mu$m in size. In response to an external stimulus the trichocysts
can be simultaneously extruded from the cell. In this process, the 
crystalline trichocysts expand in length by a factor of roughly eight.
The purpose seems to be defensive. The explosive release
of the trichocysts can push {\em Paramecium} away from a potential predator, 
giving it a chance to escape.\cite{Knoll91}

In both these examples, a high level of control is asserted over the
crystallization process in order to determine the location, size and shape of the 
crystal. The mechanical properties are also important for their function with
the virus-encapsulating crystals required to be tough in order to provide 
a protective environment, and the trichocysts able to undergo 
an irreversible expansion triggered by the presence of calcium ions.

Our final example is of the crystallization of virus particles themselves 
(Fig.\ \ref{fig:invivo}(c)),
and has perhaps more to do with an absence of a selection pressure to 
prevent crystallization (why should it bother the virus?) than a functional
purpose for the crystals. The most dramatic examples come from iridoviruses,
which have large icosahedral capsids. This type of virus was first identified
due to the iridescent colours imparted to insect larvae 
(hence the name)\cite{Williams57} due to the Bragg scattering of visible 
light by the crystals.\cite{Klug59}

\section{Glass formation}
\label{sect:glass}
There has been much work trying to understand the unusual dynamic properties
of supercooled liquids as the glass transition approached.\cite{Debenedetti01}
To do this from 
a simulation perspective, model systems have been developed 
that are robust glass-formers and show no tendency to crystallize. 
Such models also allow one to probe 
the structural determinants of a system's glass-forming ability. 
A key notion is that of `frustration',\cite{Sadoc99,Tarjus05} which is 
said to occur when the preferred local order is incompatible with global 
crystalline order. Typically, the liquid structure reflects the preferred local 
order, and so when frustration is present, significant structural
reorganization will be necessary to nucleate a crystalline phase.

To apply these ideas, one needs to be able to identify the preferred local 
order, but, because of the inherent disorder associated with a liquid, 
sometimes this can be hard to achieve directly from
analysing liquid configurations.
An alternative approach is to instead look at the structures of isolated 
clusters, since such a cluster can adopt the preferred structure
without having to mould itself to any surrounding medium.  
This approach was first applied by Charles Frank to rationalize
why small liquid metal droplets could be substantially 
supercooled.\cite{Frank52} 

\begin{figure}[t]
\begin{center}
\includegraphics[width=8.5cm]{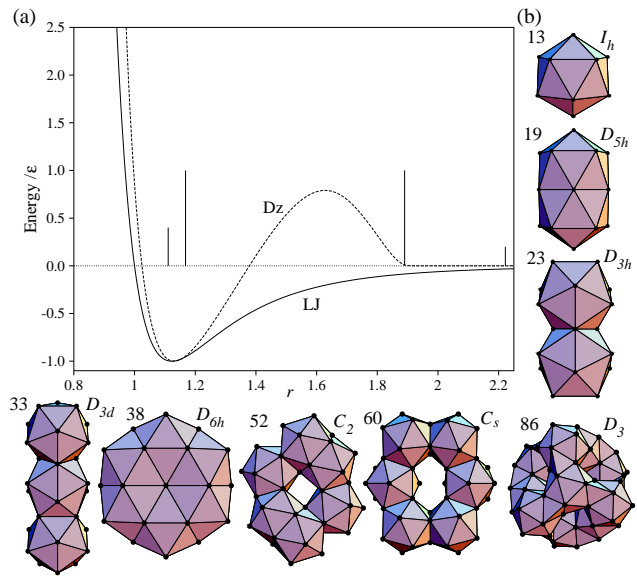}
\caption{\label{fig:Dz} (a) The Dzugutov (Dz) and Lennard-Jones (LJ) potentials
compared. The vertical lines show the positions of the pair distances
in the 13-atom icosahedron and their heights are proportional to the numbers of
pairs with that distance.
(b) Some of the particularly stable structures for Dzugutov clusters. 
Each cluster is labelled by its size and point group.}
\end{center}
\end{figure}

One of the potentials most commonly used by simulators is the Lennard-Jones 
potential.  
However, the one-component Lennard-Jones fluid is not a good 
system to look at the behaviour of supercooled liquids, 
since crystallization into a close-packed solid occurs relatively easily.
To generate a good glass-former, Dzugutov modified this potential by 
adding a barrier in the potential at roughly $\sqrt 2$ times the position
of the potential minimum (Fig.\ \ref{fig:Dz}(a)), hence energetically 
disfavouring close-packed crystals because of their octahedral 
interstices.\cite{Dzugutov92}
Instead, polytetrahedral order,\cite{NelsonS} 
and local icosahedral order in particular, is preferred. 
As can be seen in Fig.\ \ref{fig:Dz}(a), the distances 
present in the 13-atom icosahedral cluster avoid sampling this bump in 
the potential.

Even though there is a stable body-centred-cubic phase,\cite{Roth00} 
the resulting Dzugutov liquid can be easily supercooled. Interestingly,
the Dzugutov system is one of the only liquids in which an increase in
a structural length scale has been detected on increased supercooling,
in particular domains with local icosahedral order grow in
size as the temperature is decreased.\cite{Dzugutov02b} 
However, these domains are not compact, but are
ramified in structure. These results raise the questions: What are the origins
of the non-compactness, and might this shape be a generic property of 
domains of enhanced local order within supercooled liquids? 
An analysis of the structures of isolated clusters is
able to provide some answers. This type of ordering is also clearly seen in 
the clusters, but an analysis of the energetics shows that it
is a consequence of the unusual shape of the Dzugutov potential.
Therefore, this domain shape is unlikely to be universal.

The lowest-energy Dzugutov clusters
are all aggregates of interpenetrating and face-sharing icosahedra
(Fig.\ \ref{fig:Dz}(b)).\cite{Doye01a} 
As the cluster size increases first chains, then discs, then rings, and 
finally 3-dimensional porous networks of icosahedra are seen. The energetic
causes for this ordering are also clear. The bump in the potential promotes
local icosahedral order. However, icosahedral structures are inherently
strained, e.g.\ the distance between adjacent atoms on the surface of 
the regurlar 13-atom icosahedron is 5\% longer than that to the central atom
(Fig.\ \ref{fig:Dz}(a)). 
This strain, and the associated energetic penalty, grows rapidly for compact
polyicosahedral structures. The resulting non-compact icosahedral aggregates
represent a compromise that maintains the local icosahedral coordination, 
but which does not involve excessive strains.
Similar behaviour has also been seen for variants of the Dzugutov
potential.\cite{Doye03b}

A second approach used to generate a good glass-former is to 
introduce two atom types. Partly this is because
the formation of a compositionally ordered crystal can require the atoms
to diffuse significantly further to find the correct environment
than for the one-component case. However, there is more to it than this, 
since the interactions between the particles and the composition also need to 
be tuned to reach those regions of the parameter space where crystallization is
particularly difficult.\cite{Fernandez03,Fernandez04} 
The most commonly used model of this type is the Kob-Andersen
binary Lennard-Jones mixture,\cite{Kob94} i.e.\ the potential is 
\begin{equation}
V_{\rm BLJ}(r_{ij})=4 \epsilon_{\alpha\beta} \left[
             \left({\sigma_{\alpha\beta}\over r_{ij}}\right)^{12}-
             \left({\sigma_{\alpha\beta}\over r_{ij}}\right)^6
                                    \right],
\end{equation}
where $\alpha$ and $\beta$ are the atom types of atoms $i$ and $j$,
and $\epsilon_{\alpha\beta}$ and $2^{1/6}\sigma_{\alpha\beta}$
are the pair well depth and equilibrium pair separation, respectively,
for the interaction between atoms $i$ and $j$.
The Kob-Andersen parameters,
$\sigma_{AB}=0.8\,\sigma_{AA}$, $\sigma_{BB}=0.88\,\sigma_{AA}$, 
$\epsilon_{AB}=1.5\,\epsilon_{AA}$, and $\epsilon_{BB}=0.5\,\epsilon_{AA}$,
are non-additive and strongly favour mixing.
At the canonical composition A$_4$B the ground state is a
coexisting pure A face-centred-cubic crystal, and an AB CsCl-type 
crystal,\cite{Fernandez03} with low-lying layered A$_4$B crystals also 
possible.\cite{Middleton01b}
However, crystallization has never been seen in a simulation.

\begin{figure}[t]
\begin{center}
\includegraphics[width=8.5cm]{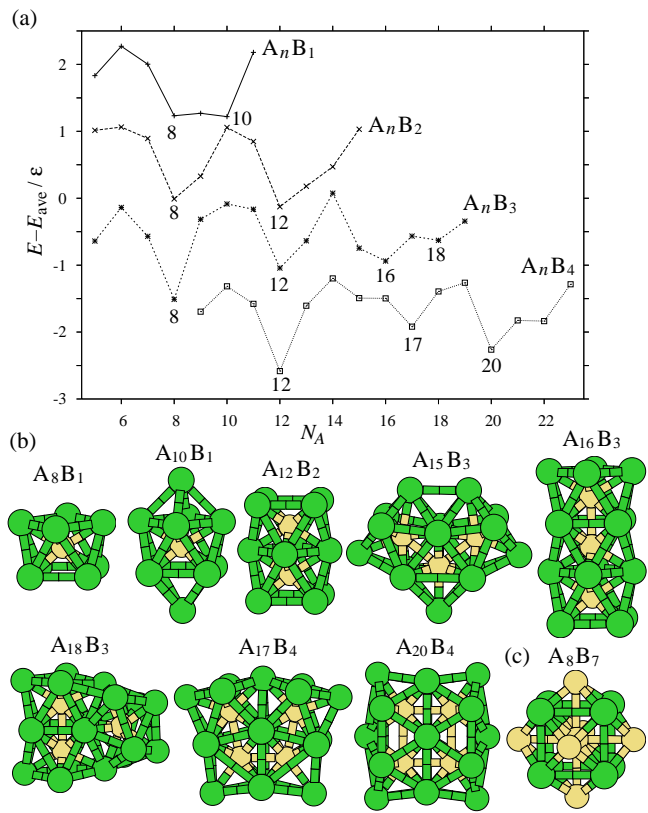}
\caption{\label{fig:BLJ} (a) The energies of the global minima for four
series of Kob-Andersen BLJ clusters with different numbers of B atoms.
To make particularly stable clusters stand out, the energies 
are measured with respect to $E_{ave}$, a fit to the energies of the clusters.
For clarity, the curves have also be  displaced with respect to each other.
(b) Some of the particularly stable structures identified in 
(a) with all B atoms completely coordinated.
(c) A particularly stable cluster at near equimolar composition.}
\end{center}
\end{figure}

Here, we use isolated clusters to analyse the preferred coordination 
environments of A atoms around B atoms, since at the glass-forming composition
B atoms are in a minority. In particular, we locate the particular
stable clusters with one, two, three and four B atoms to see how these
preferred environments can organize into larger structures 
(Fig.\ \ref{fig:BLJ}).
For coordination of a single B atom, clusters with 8, 9 and 10 atoms
have similar stability. All three structures are based on a square antiprism 
of A atoms, but with possible capping atoms over the two square faces. 
This stability of the square antiprism is also seen for A$_8$B$_2$ and 
A$_8$B$_3$, where the additional capping atoms are now B atoms. One way 
of combining such coordination environments is illustrated by A$_{12}$B$_2$
and A$_{16}$B$_3$ and involves the sharing of the square-faces of the square
antiprisms to form linear aggregates. Similar columns of square anti-prisms 
are found in the Al$_2$Cu crystal. A$_{18}$B$_3$ illustrates another 
possibility and involves the sharing of two triangular faces of A$_{12}$B$_2$
with a bicapped square antiprism. 

Another possible coordination environment for the B atoms is the tricapped 
trigonal prism. Although this structure is not stable for A$_9$B (it can 
relax to a monocapped square antiprism by a single diamond-square 
process\cite{Lipscomb}) it becomes a common environment for larger clusters. 
This is illustrated by the structures A$_{15}$B$_3$, A$_{17}$B$_4$ 
and A$_{20}$B$_4$, which have one, two and four tricapped trigonal 
prismatic environments, respectively, albeit with B atoms in some of the 
capping sites.

These results tie in well with studies of the local structure in liquid 
configurations, which have found that square antiprism, and trigonal prismatic
environments increasingly dominate the local structure in the liquid
as the temperature is decreased.\cite{Fernandez04b}

So how do these results help to rationalize the system's ability to 
avoid crystallization. Firstly, these preferred local environments are not 
present in the lowest-energy crystal structures. However, there are still 
a number of crystalline structures that are significantly lower in energy
than the liquid that do involve these 
environments.\cite{Fernandez03,Fernandez04,Fernandez04b} 
Secondly, the diversity of environments and ways that these can pack is 
likely to frustrate the formation of a uniform crystal.

In contrast to the A$_4$B system, at equimolar compositions the Kob-Andersen
BLJ model easily crystallizes into a CsCl-type structure.\cite{Fernandez03}  
Again, isolated clusters can help to understand this behaviour. 
Examining A$_8$B$_n$ clusters, we found A$_8$B$_7$ to be particularly 
stable and to exhibit the bulk crystal structure (Fig.\ \ref{fig:BLJ}(c)), 
hence showing the absence of frustration at this composition.

\section{Patchy models}
\label{sect:patchy}

In the above models with isotropic interactions, the connection between the
preferred local structure and the form and parameters of the potential
can be quite subtle.
To probe the connection between crystallizability and local structure further,
it would be useful to have a model where the local structure can be more 
directly controlled. 
To achieve this goal in a simple and flexible way, here we use `patchy' particles, 
where the particles are spherical and only interact strongly when the 
patches on adjacent particles are aligned, but we should note that other ways
of using anisotropic interactions to control the local structure have recently 
been explored\cite{Shintani06,Molinero06}.
There has been much recent interest in such patchy models from the perspective 
of protein crystallization,\cite{Sear99c,Dixit02,Kern03,Chang04,Talanquer05}
self-assembly\cite{Zhang04,Zhang06} 
(particularly into monodisperse clusters in a way similar to 
virus self-assembly\cite{Zhang04,Hagan06,Wilber06}) 
and the dynamics of supercooled liquids,\cite{deMichele06,Bianchi06} but it is 
still an area that is very much in its infancy with much to be discovered. 
Our interest is to probe how the anisotropy of the interparticle interactions 
can be used to control a system's crystallization and self-assembly behaviour,
and, in particular, through its effect on the degree of frustration.
As well as being of fundamental interest, the hope is that the results will
also be useful for understanding the crystallization behaviour of proteins
and anisotropic colloids.

\subsection{Potential}

\begin{figure}[t]
\begin{center}
\includegraphics[width=7.0cm]{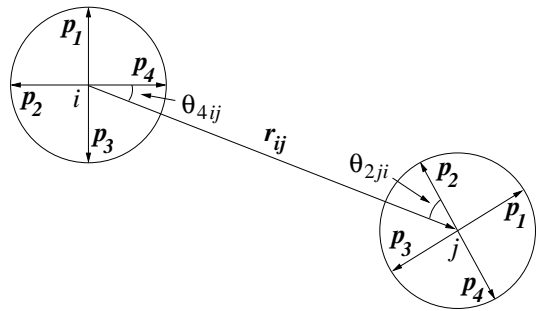}
\caption{\label{fig:potential} The geometry of the interactions between the 
model particles. In this example, there are four patches regularly arranged 
on the discs, with their directions described by the patch vectors, 
${\mathbf p_i}$. 
Patch 4 on particle $i$ interacts with patch 2 on particle $j$ because they are 
closest to the interparticle vector.}
\end{center}
\end{figure}

Here, we model the patchy particles with a single-site potential, i.e.\ each 
particle is represented by just a single site, but which has both position and 
orientation.
The potential consists of an isotropic repulsion, which 
is based on the Lennard-Jones potential
\begin{equation}
V_{\rm LJ}(r)=4\epsilon \left[ \left({\sigma_{\rm LJ}\over r}\right)^{12}-
                           \left({\sigma_{\rm LJ}\over r}\right)^6 \right]
\end{equation}
but where the attraction is modulated by an orientational dependent term, 
$V_{\rm ang}$.  Thus, the complete potential is
\begin{equation}
\label{eq:potential}
V_{ij}({\mathbf r_{ij}},{\mathbf \Omega_i},{\mathbf \Omega_j})=\left\{ 
    \begin{array}{ll}
       V_{\rm LJ}(r_{ij}) & r<\sigma_{\rm LJ} \\
       V_{\rm LJ}(r_{ij}) 
       V_{\rm ang}({\mathbf {\hat r}_{ij}},{\mathbf \Omega_i},{\mathbf \Omega_j})
                       & r\ge \sigma_{\rm LJ}, \end{array} \right.
\end{equation}
where ${\mathbf \Omega_i}$ is the orientation of particle $i$.
The patches are specified by a set of patch vectors, $\{{\mathbf p_i}\}$, 
as illustrated 
in Fig.\ \ref{fig:potential}. $V_{\rm ang}$ has the form.
\begin{equation}
V_{\rm ang}({\mathbf {\hat r}_{ij}},{\mathbf \Omega_i},{\mathbf \Omega_j})=
\exp\left(-{\theta_{k_{\rm min}ij}^2\over 2\sigma^2}\right)
\exp\left(-{\theta_{l_{\rm min}ji}^2\over 2\sigma^2}\right)
\end{equation}
where $\sigma$ is the standard deviation of the Gaussian, 
$\theta_{kij}$ is the angle between the 
patch vector $k$ on particle $i$
and the interparticle vector $\mathbf r_{ij}$, 
and $k_{\rm min}$ is the patch that minimizes the magnitude of this angle. 
Hence, only the patches on each particle that are closest to the interparticle 
axis interact with each other, and the potential is continuous as a function of 
the orientations of the particles. 
$V_{\rm ang}=1$ when two patches point directly at each other, 
but falls off as the patches deviate further from the perfect alignment.

\begin{figure*}[t]
\begin{center}
\includegraphics[width=18cm]{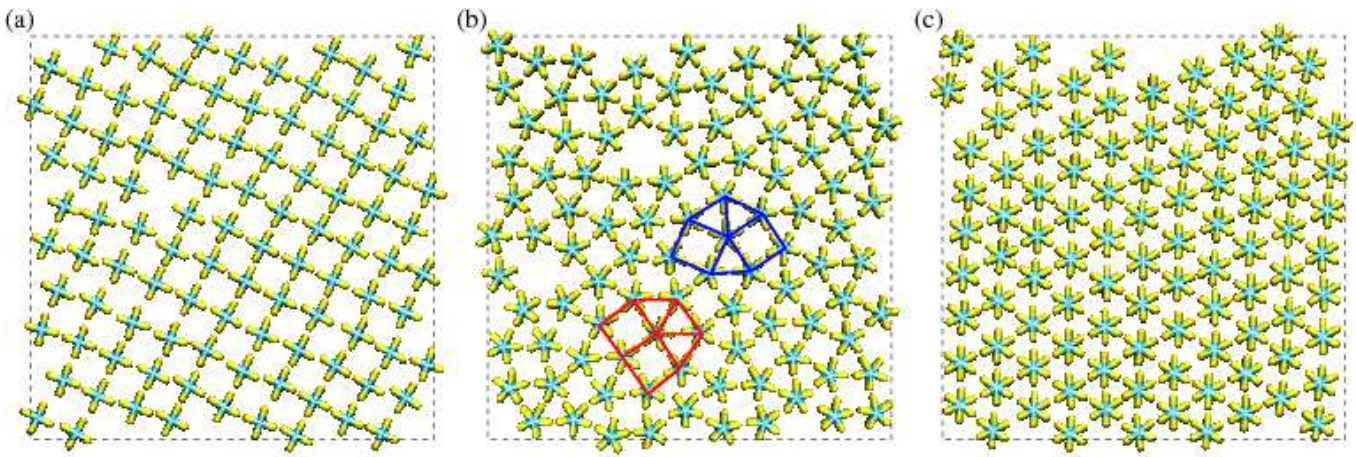}
\caption{\label{fig:2Dcool} 
Structures resulting from Monte Carlo cooling simulations for two-dimensional 
particles with (a) four (b) five and (c) six regularly spaced patches.
In each case, there are 100 particles and $\sigma=\pi/12\approx0.262$.
The pressures are (a) 0.1, (b) 0.2 and 
(c) $1.0\,\epsilon/{\sigma_{\rm LJ}}^3$.
The dashed lines show the periodic boundary conditions. 
In (a) and (c) crystallization clearly occurs. 
The crystals are not quite perfect, because unless crystallization 
occurs at the correct orientation with respect to the boundary conditions
and with the correct number of lattice points in each direction, there will 
not be the right number of particles to form a perfect crystal, and so
some defects are necessarily present.
In (b) there is no overall crystalline order, but two common motifs are 
highlighted.
}
\end{center}
\end{figure*}

One of the nice features of this single-site potential is that, given a set
of patch vectors, the potential has only one parameter $\sigma$, which 
determines the widths of the patches. 
Furthermore, as
$1/\sigma\rightarrow 0$ 
the isotropic Lennard-Jones potential is recovered.
Hence, it is possible to systematically study the behaviour of the model
as a function of the degree of anisotropy with the well-characterized 
Lennard-Jones model as one limit.

\subsection{Two-dimensional crystallization}
\label{sect:2D}
We first discuss the application of this model to crystallization in 
two dimensions to illustrate the effects of the geometrical arrangement of 
the patches, since visualization is easier than in three dimensions.
We choose to study particles with 4, 5 or 6 patches 
arranged regularly on the surface of the disks. 
Particularly interesting will be the 5-patch system as the local five-fold 
symmetry of the patches is incommensurate with global crystalline order. 

The two-dimensional Lennard-Jones reference system crystallizes 
into a close-packed crystal with a hexagonal arrangement of neighbours around 
each particle.  Unsurprisingly, for particles with a regular hexagonal 
array of patches, the anisotropy reinforces this behaviour.
Crystallization is easy (Fig.\ \ref{fig:2Dcool}(c)) and the 
close-packed crystal is lowest in energy for any combination of 
pressure and $\sigma$.

\begin{figure}[t]
\begin{center}
\includegraphics[width=8.5cm]{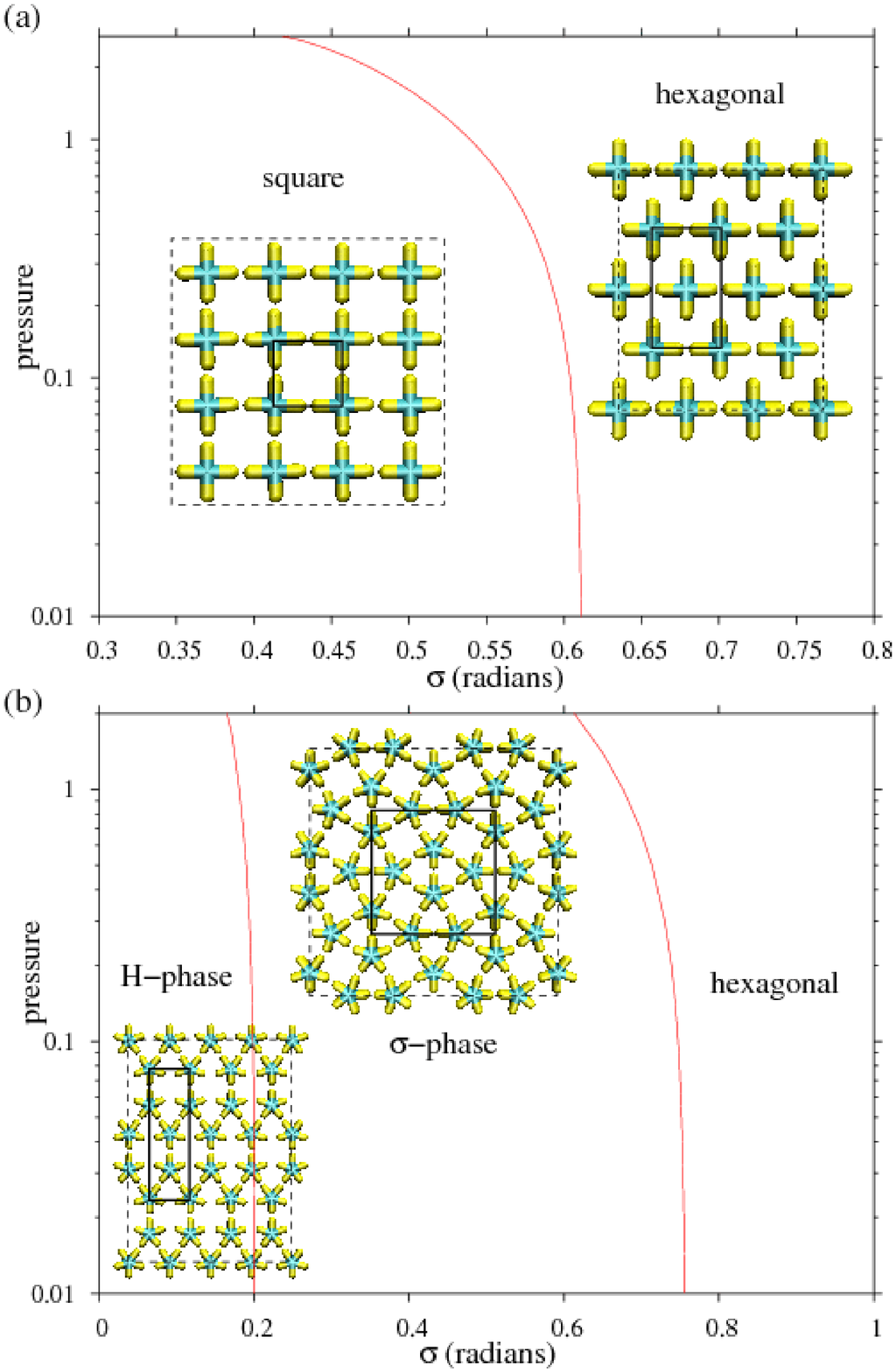}
\caption{\label{fig:2Dphase} 
Structural phase diagram showing how the ground-state structure in two 
dimensions depends on patch width and 
pressure (in units of $\epsilon/{\sigma_{\rm LJ}}^3$)
for particles with (a) four and (b) five regularly spaced
patches. The unit cells of the crystalline structures 
are depicted with thick lines, and the dashed lines show the periodic
boundary conditions. }
\end{center}
\end{figure}

In the 4-patch system there is competition between a low-density square 
crystal in which each patch points directly at a nearest neighbour, 
and a higher-density hexagonal crystal. The latter is orientationally
ordered with two opposite patches pointing at nearest neighbours, and the 
other two at next neighbours. 
As the square crystal has more pairs of nearest-neighbour patches pointing
directly at each other, it is energetically preferred when the 
patches are sufficiently narrow (Fig.\ \ref{fig:2Dphase}(a)). 
However, because of its more open structure, 
it becomes destabilised relative to the hexagonal crystal as the pressure is
increased.
Crystallization to the square crystal occurs relatively easily in 
the appropriate region of the phase diagram (Fig.\ \ref{fig:2Dcool}(a)).

For the 5-patch system the situation is more complex. No crystal phases
are possible where all the patches point directly at neighbouring particles.
For example, on cooling this system in the region of the phase diagram where
the patches prefer to be aligned, 
the resulting configuration has no overall crystalline order.
However, most of the particles have one of the two local environments
highlighted in Fig.\ \ref{fig:2Dcool}(b). From these two motifs, two crystals
can be constructed where every particle has an identical environment 
(Fig. \ref{fig:2Dphase}(b)).
In both crystals each particle has five nearest neighbours, 
and the crystals can be thought of in terms of tilings of squares and 
equilateral triangles, where every vertex is surrounded by three triangles and 
two squares. The nomenclature for
these semi-regular tilings is ($3^2.4.3.4$) and ($3^3.4^2$), and derives from
the ordering of polygons around each vertex.\cite{Grunbaum}
However, we will use the notation `$\sigma$' and `H' by analogy
to the Frank-Kasper phases of these names that can be envisaged as 
square-triangle tilings in two of their three dimensions.\cite{Shoemaker}

When the patches are sufficiently narrow, 
the H- and $\sigma$-phases
represent the best compromise between the five-fold symmetry of the particles
and crystalline order.
However, the patches cannot point directly at each other in these crystals, and
so the system is frustrated.
The mean deviation of the patches away from the nearest-neighbour 
interparticle vectors is smaller for the $\sigma$-phase ($7.2^\circ$) 
than for the H-phase ($9.6^\circ$), 
making the former lowest in energy for intermediate values of $\sigma$ at low
pressure. However, at small $\sigma$, the energetic penalty associated with 
these deviations becomes significant, and causes the H-phase to distort so that 
three of the five patches can point directly at each other. This distortion
reduces the aspect ratio of the unit cell and the triangles deviate 
significantly from the ideal equilateral geometry. No such similar distortion
is possible for the $\sigma$-phase, and so the H-phase becomes lowest
in energy at small $\sigma$ (Fig. \ref{fig:2Dphase}(b)).

The H- and $\sigma$-phase represent only two of the possible tilings of 
squares and triangles. These tilings need not be crystalline, 
and dodecagonal quasicrystalline packings are also possible.
Indeed, these quasicrystals have been seen in both metallic 
alloys\cite{Shoemaker} and macromolecular systems,\cite{Zeng05} 
under circumstances near to where crystalline square-triangle phases are stable.
Therefore, a stable quasicrystalline state remains an intriguing possibility
for the current system. 

The more dense hexagonal crystal is most stable at higher pressure, and
closer to the isotropic limit. Although positionally ordered, the crystal
is orientationally disordered, because of the incompatibility of the
five-fold symmetry of the particles and the six-fold symmetry of the crystal 
(Fig. \ref{fig:2Dphase}(b)).

An interesting comparison to the current system, which has five-fold symmetry
in the attractive interactions, is provided by a study of 
hard pentagons.\cite{Schilling05}
Although the shape of these particles does frustrate crystallization to some
extent, the lattice formed by the particle centres in the only stable 
orientationally-ordered crystalline phase, is based on a slightly distorted 
hexagonal lattice.

Also relevant is a recent study of a two-dimensional system of particles
interacting with a Lennard-Jones potential plus an anisotropic term that 
favours the formation of isolated five-fold rings of particles\cite{Shintani06}.
The presence of such order again frustrates crystallization, and, 
interestingly, on varying the strength of the anisotropic term in the potential,
the dynamic properties of the liquid, such as the fragility, can be changed 
substantially.

\subsection{Three-dimensional crystallization}
\label{sect:3D}

In our applications of this model to three dimensions, we again look at
how low-density crystalline phases can be stabilized by the patchy interactions.
The two systems that we consider are 4- and 6-patch particles 
with a regular tetrahedral and octahedral arrangement of the patches, 
respectively. 
For these systems, the stable crystalline phases at sufficiently low pressure 
and $\sigma$ would be expected to be a diamond and a simple cubic lattice, 
respectively, with each patch directly point at a neighbouring particle.
These have maximum packing fractions of 34\% and 52\% compared to 
74\% for the face-centred-cubic crystal favoured by the isotropic Lennard-Jones
potential.

\begin{figure}[t]
\begin{center}
\includegraphics[width=8.5cm]{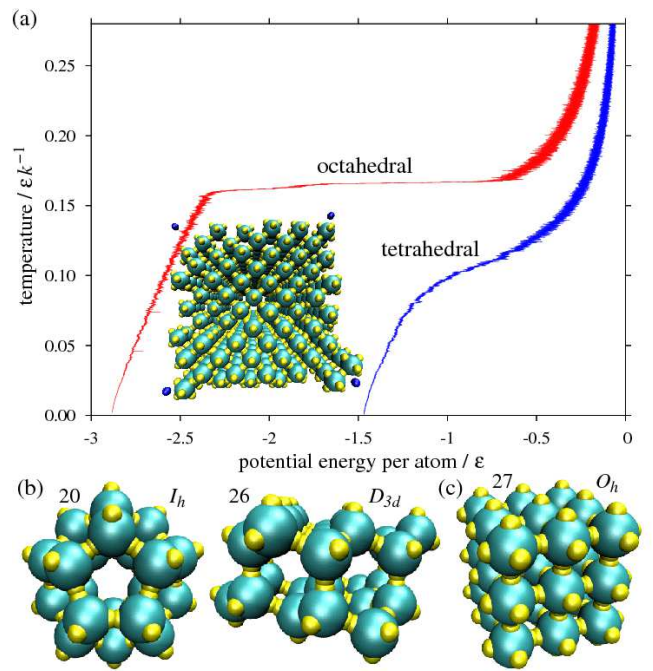}
\caption{\label{fig:3D} (a) The caloric curves for cooling the 4-patch 
tetrahedral particles and the 6-patch octahedral particles from high 
temperature. The number of particles is 512, 
the patch width $\sigma$ is 0.3 radians, and
the pressure is $0.1\,\epsilon/{\sigma_{\rm LJ}}^3$. 
The resulting simple cubic crystal for the 6-patch system is also illustrated
(the dots represent the corners of the simulation box). 
(b) and (c) Some of the global minima of isolated clusters 
for the 4- and 6-patch systems, respectively.
Each cluster is labelled by 
the number of atoms and point group.
}
\end{center}
\end{figure}

One might have expected these two systems to behave quite similarly in the
regime where the open structures are preferred, but in fact their 
crystallization behaviour is quite different. The 6-patch system
is able to crystallize easily, leading to a step-like decrease in the
energy on crystallization (Fig.\ \ref{fig:3D}(a)), 
sometimes even giving a perfect defect-free crystal. 
By contrast for the 4-patch system at best only partial crystallization
is observed, and on ordering the energy does not exhibit a step,
but instead changes continuously.

As in Section \ref{sect:glass}, examining the structure of isolated clusters
can give us some clues to the physical origins of this behaviour, because the
clusters provide a picture of the preferred local order for the system.
For both systems, when the patches are sufficiently narrow, as expected,
the lowest-energy clusters exhibit open structures where the maximum 
coordination number is equal to the number of patches. 
For the octahedral particles, for $\sigma\lesssim 0.5$ most of the 
clusters are cuboidal nanocrystals with the simple cubic structure 
(e.g.\ the $3\times 3\times 3$ cube in Fig.\ \ref{fig:3D}(c)). 
The system is unfrustrated and it is unsurprising that crystallization is 
relatively easy.

By contrast, the global minima of the tetrahedral system for 
$0.2\lesssim\sigma\lesssim 0.45$ do not exhibit the structure of 
the stable crystalline phase, but instead are based on dodecahedral cages 
(the 20-particle dodecahedron is shown in Fig.\ \ref{fig:3D}(b)).
Only for $\sigma\lesssim 0.2$ are clusters with the diamond structures lowest
in energy, e.g.\ the 26-particle structure in Fig.\ \ref{fig:3D}(b). 
In diamond the particles 
form six-fold rings, but which are puckered, whereas the dodecahedra are 
characterized by planar five-fold rings, where the 108$^\circ$ bond angles
are very close to the 109.57$^\circ$ angles between the tetrahedral
patches. The effect of this difference is small until the patches become
very narrow, and so for intermediate values of $\sigma$, the dodecahedral
clusters are preferred because they have a greater number of bonds, e.g.\
for the 20-atom dodecahedron there is only one unused patch per particle, 
whereas the particles on the surface of diamond clusters often have two
unused patches. 
Thus, the results for these clusters suggest that the tendency of the liquid
to form structures involving five-fold rings is one of the reason underlying
the much greater difficulty of crystallization in the tetrahedral system.

Furthermore, as well as this competition between five- and six-fold rings, 
in this model the diamond and hexagonal diamond (the one-component equivalent of
wurtzite) structures are practically degenerate. The structural difference 
between these two crystals is that all the six-fold rings in diamond have a 
form analogous to the chair isomer of cyclohexane, whereas in hexagonal diamond,
some of these rings are analogous to the boat form of cyclohexane. Again, 
this further variety of local structural forms is not going to aid 
crystallization.

These results are consistent with the work of 
Zhang {\em et al.} who attempted to crystallize diamond using 
similar patchy particles with the same tetrahedral patch 
geometry.\cite{Zhang06}
Crystallization to the diamond structure only reliably occurred when 
a seed crystal was introduced, or when an additional term in the potential 
was added that favoured staggered over eclipsed torsional configurations, 
and hence disfavoured hexagonal diamond.

This work on model anisotropic particles is particularly timely given the
rapid recent advances in synthesizing anisotropic 
colloids,\cite{Manoharan03,Cho05,vanBlaaderen06} and because 
one target for research in this field is to get particles with 
tetrahedral symmetry to assemble into a diamond-like crystal.\cite{Ngo06} 
Our results indicate 
that the crystallization of such tetrahedral colloids might not be
so straightforward, because of the potentially frustrating effects of the
variety of local structures possible in the liquid phase. 
However, it is not clear how an additional potential 
term similar to that used by Zhang {\em et al.} could be introduced into the 
intercolloidal potential to alleviate this.  Instead, an alternative strategy 
would be to use a binary system of oppositely-charged tetrahedral colloids, 
as this would penalize the formation of rings with an odd number of particles. 
In particular, this change would reduce the tendency for pentagons to form, 
hence removing some of the frustration and making crystallization more likely.

An interesting comparison to the present results are provided
by a recent study that took the Stillinger-Weber silicon 
potential\cite{StillW85} and varied the strength of the 
anisotropic 3-body term, finding that it had a significant effect on the 
system's glass-forming ability\cite{Molinero06}.
For the original silicon potential crystallization into a diamond structure
occurred relatively easily, but as the strength of the 3-body term is decreased
the system becomes a good glass-former in the region of parameter space 
where there is a crossover in the stability of the diamond and 
body-centred-cubic crystals, and it has been suggested that this is partially 
due to the structural dissimilarity between the liquid and the possible 
crystals. Similarly, it would be interesting to look at how varying the patch
width affected the dynamics of crystallization for our systems.

\section{Conclusions}
\label{sect:conc}
Through the examples of crystallization (and its absence) considered in this
paper, we hope to have shown the subtle interplay between the interparticle
potential, the preferred local structure and the kinetics of crystallization.
In particular, if the interactions favour a local structure that is 
incompatible with the global crystalline order, or almost equally favour 
a variety of different local environments, then crystallization is likely 
to be frustrated. Furthermore, we have provided further empirical 
support that examining the structures of isolated clusters, as first done by
Frank,\cite{Frank52} can potentially provide a clearer picture of the 
preferred local order, and hence the possible presence of frustration. 

An important question concerning the type of structural approach advocated 
here is how it relates to other approaches to understanding the kinetics 
of crystallization, such as classical nucleation theory. In the latter,
one of the key parameters in determining the ease of nucleation is the
surface free energy associated with the crystal-liquid interface. The 
connection to the current approach is that this interfacial
free energy is particularly sensitive to the degree of structural
dissimilarity between the crystal and the liquid, i.e.\ 
the greater the structurally dissimilarity, the greater the interfacial free 
energy, and hence the harder crystallization becomes. 

In such instances where the liquid and crystal are structurally very different,
one way of circumventing the large free-energy barriers to direct nucleation
of the crystal is to undergo a two-step nucleation process, i.e.\ first
nucleate a metastable phase, and then nucleate the final crystal form from 
within that phase\cite{Ostwald,tenWolde99}. 
For example, for the Stillinger-Weber silicon potential
mentioned earlier, crystallization to the diamond structure only occurs 
at temperatures below that for a liquid-liquid phase transition, which leads
to the nucleation of a lower-density liquid phase that more closely resembles
the solid\cite{Molinero06}.

The patchy models introduced in Section \ref{sect:patchy} are particular
useful for studying crystallization, 
since the geometry of the patches allows the system's structural propensities 
to be directly controlled in a simple and flexible manner. 
One of our original intentions for these models was to also use them to 
illuminate aspects of the crystallization of proteins. 
Therefore, a key question is how relevant these models are to real proteins.
Clearly, the interactions between proteins are strongly anisotropic in 
character,
i.e.\ in order for two proteins (in their native state) 
to come together, the regions of their respective surfaces that come into 
contact, the ``patches'' if you like, must be correctly aligned and oriented.
Furthermore, the patchy models, like proteins, are also able to form 
relatively low-density crystals. 
However, actual interprotein interactions are much more 
complex---proteins are anisotropic in shape, their surfaces are very 
heterogeneous, and the interactions can depend sensitively on solution 
conditions, such as pH and the concentrations of other ions.\cite{Allahyarov02}
Thus, how much of this complexity one needs to capture in order to reproduce 
or gain insight into the crystallization of proteins is an open, but
important, question.  

The examples described in Section \ref{sect:pxtal} illustrate the exquisite 
control that biological systems are able to exert over the interaction
properties of proteins, in particular 
their ability to avoid native state aggregation or crystallization in the 
dense cellular environment, and the ability of particular proteins
to form into complex crystalline assemblies. There are some indications of 
how this control is achieved, e.g.\ the role of lysine residues in preventing
unwanted interactions, but there is much still to be learnt.
Such knowledge will be particularly important if rational methods 
to make protein crystallization {\it in vitro} easier 
are to be developed 
that aim to overcome the evolutionary selection of the 
surface properties of proteins to prevent native state aggregation.

In some ways, our patchy particles are probably somewhat closer to some of 
the anisotropic colloids now beginning to be 
generated.\cite{Manoharan03,Cho05,vanBlaaderen06} Therefore, it will be
particularly interesting, once it becomes possible to generate such colloids in
sufficient number and quality, and with their surfaces appropriately
functionalized, to be able to probe their phase behaviour and 
phase transformation kinetics. Hopefully, some of the insights gained from our 
patchy models will help to guide the experimentalists in their quest to 
create crystals with useful photonic properties.

\section*{Acknowledgements}
The authors are grateful to the Royal Society, the Ram\'on Areces Foundation 
and the Engineering and Physical Sciences Research Council for financial 
support.

\end{document}